\documentclass[twocolumn,prb,amsmath,amssymb,amsfonts,superscriptaddress,floatfix,aps,showpacs,notitlepage]{revtex4-1}

\usepackage{graphicx}
\usepackage{epstopdf}
\usepackage{dcolumn}
\usepackage{bm}
\usepackage{rotating} 
\usepackage{color}
\usepackage{subfigure}
\usepackage{natbib}
\usepackage{enumitem}
\usepackage[Symbol]{upgreek}
\usepackage[symbol*]{footmisc}
\usepackage{lipsum}
\usepackage{amsmath}                    
\usepackage{amssymb}                    

\usepackage{textcomp}					

\usepackage[version=3]{mhchem}
\usepackage{units}
\usepackage{verbatim}
\usepackage{url}
\usepackage[utf8]{inputenc}
\usepackage[T1]{fontenc}

\begin{document}
	
\title{Magnetic properties of Co doped Nb clusters}

\newcommand{\imm}{Radboud University, Institute for Molecules and
		Materials, Heyendaalseweg 135, 6525 AJ Nijmegen, Netherlands}
	
\newcommand{\felix}{Radboud University, Institute for Molecules and
		Materials, FELIX Laboratory, Toernooiveld 7c, 6525 ED Nijmegen,
		Netherlands}
	
\author{A. Diaz-Bachs}
\affiliation{\imm} 
	
\author{L. Peters}
\affiliation{\imm} 
	
\author{R. Logemann}
\affiliation{\imm} 
	
\author{V. Chernyy}
\affiliation{\imm}
	
\author{J. M. Bakker}
\affiliation{\felix}
	
\author{M. I. Katsnelson}
\affiliation{\imm}
	
\author{A. Kirilyuk}
\email{a.kirilyuk@science.ru.nl}
\affiliation{\imm}

\date{\today}
	
\begin{abstract}
	From magnetic deflection experiments on isolated Co doped Nb clusters we made the interesting observation of some clusters being magnetic, while others appear to be non-magnetic. There are in principle two explanations for this behavior. Either the local moment at the Co site is completely quenched or it is screened by the delocalized electrons of the cluster, i.e. the Kondo effect. In order to reveal the physical origin, we conducted a combined theoretical and experimental investigation. First, we established the ground state geometry of the clusters by comparing the experimental vibrational spectra with those obtained from a density functional theory study. Then, we performed an analyses based on the Anderson impurity model. It appears that the non-magnetic clusters are due to a complete quenching of the local Co moment and not due to the Kondo effect. In addition, the magnetic behavior of the clusters can be understood from an inspection of their electronic structure. Here magnetism is favored when the effective hybridization around the chemical potential is small, while the absence of magnetism is signalled by a large effective hybridization around the chemical potential.
\end{abstract}
	
	\pacs{Valid PACS appear here}
	\keywords{niobium clusters, cobalt, magnetic moments, Anderson impurity model}
	\maketitle
	
\section{Introduction}
Electronic correlations constitute the basis of condensed matter physics and are responsible for the enormous wealth of phenomena found in solids, such as (high-$T_{c}$) superconductivity\cite{Dagotto1994}, charge- and spin-ordering\cite{Tranquada1995} and fluctuations\cite{Yamada1998}, colossal magnetoresistance\cite{Fontcuberta1996}, metal-insulator transition\cite{Zylbersztejn1975}, half-metallicity\cite{Katsnelson2008}, quantum Hall effect\cite{Willet1987}, heavy fermion behavior\cite{Stewart1984}, etc. Reducing the size, however, leads to an extreme sensitivity of these properties to the atomic arrangement, shape, and the effects of the environment. The understanding and control of these size-driven processes is therefore crucial to maintain the pace of developments in nanoscience.
	
In this miniaturization trend, the ultimate limit is represented by atomic clusters. Such clusters are particles composed of a countable number of atoms, from the diatomic limit up to some thousands or tens of thousands of atoms. Quantum confinement effects entirely govern the behavior of matter in this size regime. The discretized electronic levels lead to sudden changes of the cluster properties, for example when changing the cluster size on an atom-by-atom basis. In the semiconductor technology there is already interest in systems with discrete energy spectra, for example quantum wells\cite{Konig2007} and quantum dots\cite{Michalet2005}. 
	
Obviously the consideration of doped instead of pure clusters offers an even broader playground for technological applications. However, doped clusters are also very interesting from a fundamental point of view. For example, it is well known that already for a single magnetic impurity in a non-magnetic metallic host interesting phenomena like Friedel oscillations\cite{Cheianov2006} and the Kondo effect\cite{Kondo1964} can occur. How or would such effects be present in clusters? Furthermore, the case of a single magnetic impurity embedded in a discrete host like a cluster offers a sensitive probe of studying the dependence of the local magnetic moment on the details of the discrete energy spectrum. This could lead to valuable insight in quenching and/or Kondo screening mechanisms. More precisely, the formation of the atomic magnetic moment is trivially described by the Hund's rules in the case of an isolated atom, but this process is far from trivial in the case of an atom embedded in an interacting host. 
	
Recently, the magnetic moment of a single magnetic impurity in a discrete host was investigated by means of the Anderson impurity model.~\cite{Lau2014} One of the things found, was that on average the local moment grows with increasing host band gap (HOMO-LUMO gap). Here on average should be understood as the local moment averaged over a number of random configurations of the discrete host energy levels for a fixed host band gap. 
Then, based on this investigation of the Anderson impurity model, the experimentally observed magnetic moments of Cr doped Au clusters were successfully explained.~\cite{Lau2015} For example, it was found that the size of the measured local moment follows the trend of the calculated band gap of the host. 
	
In this work we present a comprehensive study of the mechanisms governing the formation of magnetic moments in Co doped Nb clusters. From magnetic deflection experiments we made the interesting observation that some clusters are strongly magnetic, while others are completely non-magnetic. Note that in contrast for the Cr doped Au clusters all measured clusters were found to be magnetic. There are two possibilities for the absence of magnetism in the Nb$_{x}$Co clusters. Either the local Co moment is completely quenched or it is screened by the delocalized electrons of the cluster, i.e. the Kondo effect. From the theoretical perspective, the difficulty in explaining the observed magnetic behavior is in the treatment of the electronic correlations. Since it is not clear from the beginning whether correlations effects are weak, intermediate or strong, it is difficult to decide which theoretical approach is suitable. One could expect correlations to be stronger in small clusters than in their bulk counterparts due to a stronger localization of the wave-functions. On the other hand, for the clusters less screening channels are present, which could lead to an almost constant Coulomb interaction throughout the cluster~\cite{fexoylars}. This would render correlations effects to be unimportant. 
	
As mentioned, a priori the importance of correlations effects is not known for Nb$_{x}$Co clusters. Therefore, we conducted a combined theoretical and experimental investigation. First, we made a comparison of the experimental vibrational spectra with those obtained from a density functional theory (DFT) study. This serves two purposes. It provides the ground state geometry of the clusters. Further, due to the dependence of the vibrational spectrum on the magnetic moment, the performance of DFT in predicting the magnetic moments can be investigated. Then, in order to obtain a physical understanding of the experimentally observed magnetic behavior, we performed an analyses based on the Anderson impurity model. From this analyses it is observed that the absence of a magnetic moment is due to a complete quenching of the Co moment and not the Kondo effect. In addition, the magnetic behavior of the Nb$_{x}$Co clusters can be understood from an inspection of their electronic structure. Here magnetism is favored when the effective hybridization around the chemical potential is small, while the absence of magnetism is signaled by a large effective hybridization around the chemical potential.     
	
Co doped and pure Nb clusters have already been the topic of interest in earlier works. One of the most relevant experimental works on pure Nb clusters is the electric deflection experiment, which showed that cold clusters may attain an anomalous component with very large electric dipole moments.~\cite{Moro2003} In contrast, the room-temperature measurements showed normal metallic polarizabilities. Further, magnetic deflection experiments on pure Nb clusters showed that at very low temperatures the clusters with an odd number of atoms deflect due to a single unpaired spin that is uncoupled from the cluster lattice. In contrast, at high temperatures deflections do not take place, as in the cluster the unpaired spin becomes coupled to the lattice.~\cite{Moro2004} Far-infrared absorption spectra of small neutral and cationic Nb clusters combined with DFT calculations have revealed their geometries.~\cite{Fielicke2007} Compared to pure Nb clusters, not many works focused on Co doped Nb clusters. Experimentally an anion photoelectron spectroscopy study is performed, which showed that the addition of the Co atom for small Nb clusters induces bulk-like behavior, i.e. closing of the band gap.~\cite{Pramann2003} From the theoretical side a computational study based on DFT addressed the geometric and magnetic properties finding that Nb$_{7}$Co has no net magnetic moment, which means that the 6 $\mu_{B}$ coming from the Co atom is completely quenched.~\cite{Li2014} The disadvantage of this purely theoretical work is the lack of an experimental confirmation, which is another reason for conducting a combined experimental and theoretical study.
	
The rest of this paper is organized as follows. In Section~\ref{magdefl} we first present our magnetic deflection experiments. Then, in Section~\ref{secvib} the experimental vibrational spectra are compared with those obtained from density functional theory calculations. Based on the ground state geometries obtained from this comparison, we perform a discussion based on the Anderson impurity model in Section~\ref{theoraim} to address the presence or absence of magnetic moments in Nb$_{x}$Co clusters. Finally, in Section~\ref{concl} we present our conclusions.  
\section{Magnetic deflection experiments}
	\label{magdefl}
\subsection{Stern-Gerlach setup}
The magnetic moments of the Nb$_{x}$Co clusters were obtained by means of a Stern-Gerlach setup.~\cite{Stern1924} This setup consists mainly of three parts: the source, the magnet and the position sensitive time of flight mass spectrometer (PSTOFMS). The source is of Milani-de Heer type.~\cite{deHeer1990} The clusters are produced in the source chamber by ablation of a Nb$_x$Co$_y$ (x = 95, y = 5~\%) rod due to a Nd:YAG laser producing 532~nm light. More precisely, this laser is focused on the rod, which is inside a cavity of a tuneable volume. The cavity is connected to a pulsed valve, responsible to introduce pulses of helium, which is the carrier gas, i.e. it is responsible for the transport of the clusters across the setup. The cavity is also coupled to a nozzle. Due to a pressure gradient across the nozzle, the clusters expand supersonically. The actual creation and cooling of the clusters takes place inside the cavity. In our setup the source can be cooled down to 20~K due to a cold head. Once the cluster beam has left the cavity, it crosses a conical skimmer of 1~mm width. After the clusters are skimmed they reach a chopper, which has two purposes: cluster selection and measurement of their velocity. Then, after the chopper there are two slits to narrow the beam in both the horizontal and vertical direction. After the slits, the cluster beam reaches the magnet, i.e. a 2 wire Rabi design electromagnet.~\cite{Rabi1934} The magnet produces an inhomogeneous magnetic field that can reach a maximum strength of 2.4~T and gradient of 350~T/mm. The spins of the cluster are aligned by the magnetic field, while the cluster is deflected due to the gradient in the field. For the calibration of the magnet aluminium atoms were chosen, since they are easy to produce and their magnetic properties are well known ($\mu_B$=1/3$\mu_B$, J=1/2, m$_{J}$=$\pm$1/2).
	
After the magnet the clusters have to travel 1~m before they reach the PSTOFMS. In order to detect the clusters, they are ionized by an excimer laser producing an ultraviolet beam of 193~nm. The ionized clusters can then be directed by the electric fields of the PSTOFMS plates towards the micro channel plate (MCP) where they are detected. After the detection of the cluster, its time-of-flight is known. This time-of-flight linearly depends on the deflection of the cluster, where the proportionality constant is obtained from another calibration. For this calibration, a narrow slit is placed in the path of the excimer beam. Then, by moving the slit, the time-of-flight can be determined for each corresponding slit position, i.e. the position where the clusters will be ionized, describing the correlation between time-of-flight and deflection. Since the determined proportionality constant is known to scale as the square root of the mass, it is only necessary to perform the calibration for one specific cluster size. 
	
From the measurement of the deflection $x$ via the time-of-flight, the mass $m$ of the cluster and its velocity $v$ by means of the chopper, the average magnetic moment is determined from
\begin{equation}
		\begin{aligned}
			\langle\mu\rangle =\frac
			{xmv\textsuperscript{2}}
			{KB}.
			\label{eqavm}
		\end{aligned}
\end{equation}
Here $B$ is the magnetic field strength and $K$ is a constant that depends on the setup. This constant includes the gradient of the magnet, which is determined from the calibration by the aluminium atoms. 
	
During the measurements we observed three different deflection (or equivalently average magnetic moment via Eq.~\ref{eqavm}) profiles, see Fig.~\ref{figdefl} (a), (b) and (c). The method we used to obtain the actual value of the deflection, depends on the observed deflection profile. In case of a deflection profile centered at zero (Fig.~\ref{figdefl} (b)) we used a Gaussian fit, where the position of the peak of the Gaussian corresponds to the 'deflection'. For the double sided deflection profile (Fig.~\ref{figdefl} (c)) three peaks can be observed. Here the peak at 0 is due to the spin-relaxation by the coupling to the lattice, which can be in principle reduced by increasing the carrier gas pressure. Therefore, the deflection should be determined from the peaks away from zero. For this purpose we used three Gaussians to fit the profile, where the deflection is determined from the peak position of the Gaussians not located at zero. Finally, for single sided deflection (Fig.~\ref{figdefl} (a)) the profile is in general asymmetric, which makes a fit by a Gaussian inappropriate. In this case we take the position of the average of the peak as the deflection.  
	
In Eq.~\ref{eqavm}, the quantity $\langle\mu\rangle$ corresponds to the measured average magnetic moment of the clusters. However, not for all the three observed deflection profiles, this corresponds to the actual magnetic moment of the cluster. While it does for no deflection and double sided deflection, it does not for single sided deflection. Obviously, no deflection corresponds to a non-magnetic cluster. Double sided deflection corresponds to atomic like clusters, i.e. clusters where the magnetic moment can freely rotate. In contrast, single sided deflection corresponds to superparamagnetic clusters with the magnetic moment coupled to the lattice. Then, due to the presence of a finite temperature, not all clusters have their magnetic moment aligned with the magnetic field. Therefore, the measured average magnetic moment needs to be related to the actual magnetic moment of the cluster. For isolated clusters this is typically done by the Langevin-Debye function. In the limit of a small magnetic field this leads to the following relation  
\begin{equation}
		\begin{aligned}
			M =\frac{1}{3}\frac{\langle\mu\rangle^{2}B}{k_{B}T},
			\label{eqclusm}
		\end{aligned}
\end{equation}
where $M$ is the magnetic moment of the cluster, $T$ the temperature and $k_{B}$ the constant of Boltzman. 
	

\subsection{Results: Magnetic moments}
\begin{figure}[tb]
		\centering
		\includegraphics[width=1.0\linewidth]{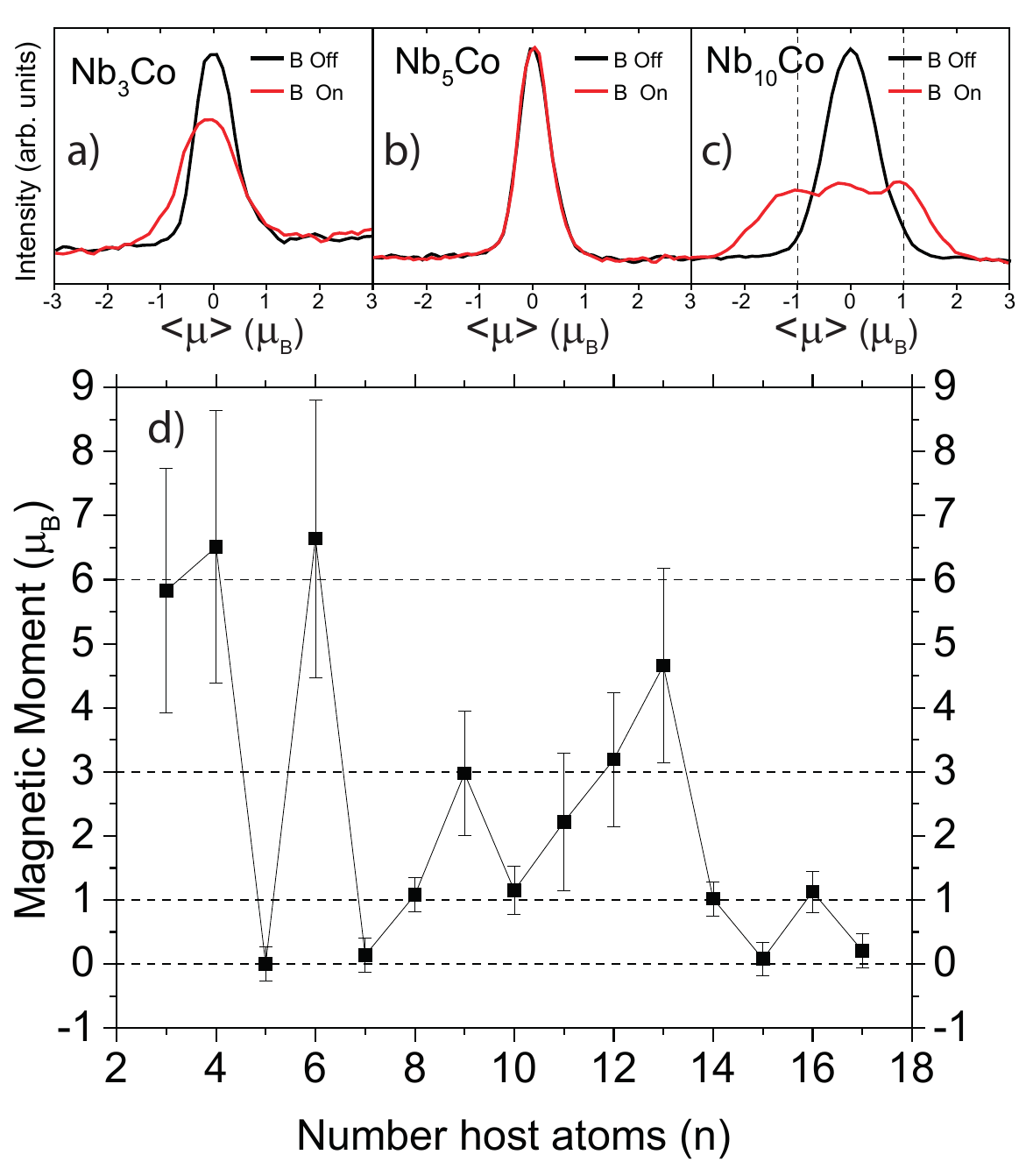}
		\caption{The top three figures contain the different deflection profiles observed for the Co doped Nb clusters. (a) Single sided deflection profile for Nb$_{3}$Co, which indicates a superparamagnetic cluster. (b) Profile of Nb$_{5}$Co showing no deflection, which corresponds to a non-magnetic cluster. (c) Two sided deflection profile for Nb$_{10}$Co, which refers to an atomic-like cluster. All deflection profiles were measured at a temperature of 25K. Further, the black and red lines correspond to the situations without and with a 2.4~T magnetic field. The only exception is in (a), where the magnetic field was 1~T. For the x-axis of (a), (b) and (c) the deflection is converted to the averaged magnetic moment via Eq.~\ref{eqavm}. In the bottom figure indicated with (d), the magnetic moment as function of cluster size is presented with $n$ corresponding to the number of host (Nb) atoms. Here the error bars are computed from the uncertainty in the velocity of the cluster and the magnetic field.}
		\label{figdefl}
\end{figure}

The results of the Stern-Gerlach experiments performed on the Co doped Nb clusters at a temperature of 25~K are presented in Fig.~\ref{figdefl}. Here Fig~\ref{figdefl} (a), (b) and (c) correspond to the three typical deflection profiles that were observed, and in Fig.~\ref{figdefl} (d) the measured magnetic moments in $\mu_{B}$ are presented as function of cluster size with $n$ corresponding to the number of host (Nb) atoms. The black and red lines in the deflection profiles correspond respectively to the situation without and with a magnetic field. The applied magnetic field was 2.4~T except for Nb$_{3}$Co, where 1~T was used.
	
Fig.~\ref{figdefl} (a) for Nb$_{3}$Co shows a typical single sided deflection profile indicating superparamagnetic behavior. Other clusters showing a single sided deflection profile were Nb$_{4}$Co, Nb$_{6}$Co, Nb$_{9}$Co, Nb$_{11}$Co, Nb$_{12}$Co and Nb$_{13}$Co. Then, Fig.~\ref{figdefl} (b) for Nb$_{5}$Co presents the situation with no deflection, which corresponds to a non-magnetic cluster. The other clusters showing no deflection were Nb$_{7}$Co, Nb$_{15}$Co and Nb$_{17}$Co. The last observed deflection profile is depicted in Fig.~\ref{figdefl} (c), where for Nb$_{10}$Co an example of a two sided deflection is given, which refers to an atomic-like cluster. This profile is characterized by 3 peaks, 2 peaks at $\pm$1$\mu_B$ and an additional peak at 0$\mu_B$. Two sided deflection was also observed for all clusters containing an odd number of atoms (i.e. with $n$ even) and with $n>=14$.
	
From the magnetic moments as function of cluster size presented in Fig.~\ref{figdefl} (d) it seems that the clusters can be divided into two regions. For clusters with $n>=14$ the magnetic to non-magnetic behavior appears to be exactly determined by having an odd or even number of atoms in the cluster. An odd number of atoms in the cluster corresponds to the situation of at least one unpaired electron and thus at least a moment of 1~$\mu_{B}$. For an even number of atoms, all the electrons can be paired. Note that the magnetic behavior of pure Nb clusters was indeed explained in this way.~\cite{Moro2004} Then, there is the regime of clusters with  $n<14$, where the magnetic behavior clearly cannot be explained due the presence or absence of a single unpaired electron. In this region strong fluctuations in the magnetic moment can be observed by just adding or removing a single Nb atom. For example, Nb$_{4}$Co is strongly magnetic, while Nb$_{5}$Co is completely non-magnetic. Then, again adding just one Nb atom leads to Nb$_{6}$Co which is again strongly magnetic. On the other hand Nb$_{7}$Co is again non-magnetic.  
	
It can also be observed that there is no cluster with a magnetic moment larger than that of an isolated Co atom. An isolated Co atom has 7 3$d$ electrons leading to a total moment of 6$\mu_{B}$, where both the spin and orbital moment contribute 3$\mu_{B}$. This indicates that the Co atom is not very effective in inducing magnetic moments in the Nb host. Further, it is interesting that Nb$_{3}$Co, Nb$_{4}$Co and Nb$_{6}$Co have a magnetic moment very close to that of an isolated Co atom. Assuming that all this magnetic moment is at the Co site, would mean that we have a situation where both the spin and orbital moment are almost unquenched.

\section{Vibrational spectra: geometric and magnetic structure}
	\label{secvib}
In this section we perform a comparison of the experimental vibrational spectra with those obtained from a DFT study. This serves two purposes. First, due to the dependence of the vibrational spectrum on the magnetic moment, the performance of DFT in predicting the magnetic moments can be investigated. Second, it provides the ground state geometry of the clusters. These ground state geometries are required as an input in Section~\ref{theoraim} to obtain a physical understanding of the observed magnetic behavior in Section~\ref{magdefl}. 
	
\subsection{Experimental details}
In order to record the vibrational spectra we coupled our cluster setup to the Free Electron Laser For Intra Cavity Experiments (FELICE). Below a brief description of the experimental setup is given and for more details the reader is referred to Ref.~\onlinecite{Jalink2015}. The clusters are produced in an ablation-type cluster source in a growth channel filled by a helium carrier gas prior to ablation of a Nb$_x$Co$_y$ (x = 95, y = 5 \%) rod by a Nd:YAG laser (532~nm). The temperature of the extension tube, which is attached to the cluster source for better cluster thermalization, is 77~K. After expansion in the source chamber, the mixture of clusters and carrier gas is skimmed. This results in the formation of a molecular beam that is shaped by a slit with a width of 0.45~mm . The interaction between the IR light and the molecular beam takes place in the center of the extraction region of the REToF mass spectrometer with a 35$^\circ$ angle between the two beams. The clusters are ionized by a frequency doubled dye laser with a photon energy of 5.4~eV entering the extraction region at a $\sim$90$^\circ$ angle with respect to the cluster beam. Then, the IR pulse energies calculated inside the FELICE cavity range between 0.2 and 0.6~J over the IR scans. The IR pulse consists of a 9~$\mu$s long train of micro-pulses with 1~ns time delay between them. The experiment operates at twice the FELICE frequency which allows to record a signal with (I$_{IR+UV}(\omega)$) and without (I$_{UV}$) IR radiation in a shot-to-shot manner. The experimental IR curves are presented in terms of gain spectra (G($\omega$)) calculated as
\begin{equation}
		\label{GainEquation}
		\begin{aligned}
			G(\omega) =\frac
			{I_{\textsc{ir+uv}}(\omega)-I_{\textsc{uv}}}
			{I_{\textsc{uv}}},
		\end{aligned}
\end{equation}
at an IR frequency $\omega$, and are IR power corrected.
	
\subsection{Computational details}
For the calculation of the vibrational spectra we employed the DFT implementation of the Vienna ab initio simulation package (VASP).~\cite{Kresse1996} The projector augmented wave (PAW) method\cite{Blochl1994,Kresse1999} in combination with the Perdew, Burke, and Ernzerhof (PBE) functional is used.~\cite{Perdew1996}
For all cluster sizes we searched for the lowest-energy geometries by using a genetic algorithm (GA)\cite{Johnston2003} in combination with DFT. The details of the used method can be found in Ref.~\onlinecite{Logemann2015}.
In addition, we also considered conformations previously reported in the literature (\ce{Nb3Co}, \ce{Nb4Co}, \ce{Nb5Co}, \ce{Nb6Co}, \ce{Nb7Co})\cite{Li2014} and re-optimized the mentioned structures. For some clusters the GA results were equal to those already found literature, while for other clusters additional geometries lower in energy were obtained (see Sections~\ref{nb3co}-\ref{nb9co} for details). Further, for the PAWs an energy cutoff of $4293$~eV is used. All forces were minimized below $10^{-3}$~eV/\r{A}. In order to eliminate inter-cluster interactions, the clusters were placed in a cubic periodic box with 16~\r{A} dimensions. For the calculations, a single \textbf{k} point ($\Gamma$) is used.
	
\subsection{Results: Geometric and magnetic structure}
Below the calculated geometries of the clusters are presented by a stick model, i.e. the clusters are presented by connected sticks. Here green correspond to \ce{Nb} and gold to \ce{Co}. Further, to facilitate the comparison of the experimental and calculated results, the experimental spectra are shown with black squares accompanied by a three-point adjacent average (blue line). The gray dashed line indicates the IR power corrected experimental spectrum. The calculated harmonic vibrational frequencies (vertical sticks) are convoluted with a 15~$\text{cm}^{-1}$ FWHM Gaussian line shape function. All frequencies for the structures presented in this work are unscaled and the energies contain the zero-point vibrational energies (ZPVE). Finally, the insets of the figures below show the energy as a function of magnetization for the presented geometries with respect to that of the ground state. 
	
\subsubsection{\ce{Nb3Co}}
	\label{nb3co}
	For \ce{Nb3Co} a trigonal pyramid is found with three different magnetic states. Here the Nb-Nb and Nb-Co distances differ slightly between the magnetic configurations.
\begin{figure}[h!]
		\centering
		\includegraphics[width=1.0\linewidth]{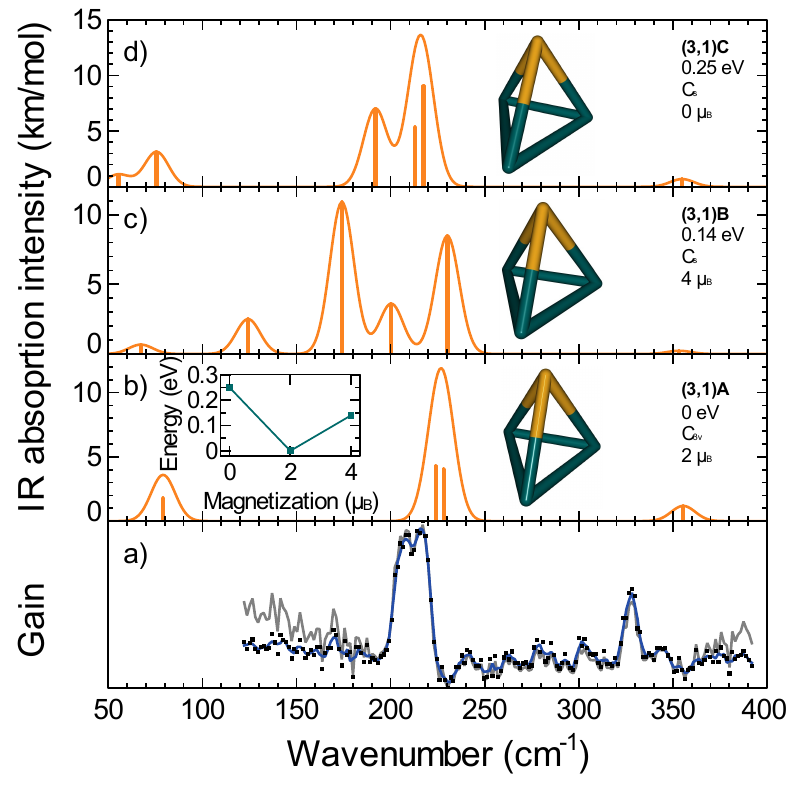}
		\caption{Experimental (panel (a), squares) and calculated ((b)-(d)) IR spectra of \ce{Nb3Co}. The blue line is three-point adjacent average of the experimental data. The gray dashed line indicates the IR power corrected spectrum. 
			The calculated discrete vibrational frequencies (orange vertical lines) are convoluted
			with a 15~$\text{cm}^{-1}$ FWHM Gaussian line shape function (orange). For the geometries green and gold are used for Nb and Co respectively. The inset graph shows the energy as function of the magnetization for the different magnetic states. }
		\label{fignb3}
\end{figure}
In Fig.~\ref{fignb3} (b)-(d) the corresponding geometries are shown. The magnetic $M=2$~$\mu_{\text{B}}$ (3,1)A geometry is lowest in energy, whereas (3,1)B and (3,1)C are 0.14~eV and 0.25~eV higher in energy respectively.
Note that geometry (3,1)A has been reported previously also as the ground state in Ref.~\onlinecite{Li2014}. The symmetry point group depends on the magnetization, with $C_{3v}$ for (3,1)A and $C_{s}$ for (3,1)B and (3,1)C. This difference in symmetry clearly results in significant differences in the vibrational spectra. 
	
Fig.~\ref{fignb3} shows that the vibrational spectrum of (3,1)A with modes at 224, 228 and 356~$\text{cm}^{-1}$ provides the best match to the experimental modes at 212 and 328~$\text{cm}^{-1}$ and also resolves the internal structure of the band at 212~$\text{cm}^{-1}$. The vibrational spectra of (3,1)B and (3,1)C contain vibrational modes in the range 125-220~$\text{cm}^{-1}$ where no experimental modes are observed. Therefore, geometry (3,1)A in the $M=2$~$\mu_{\text{B}}$ state corresponds to the ground state of \ce{Nb3Co}. 
	
\subsubsection{\ce{Nb4Co}}
\begin{figure}[htb]
		\centering
		\includegraphics[width=1.0\linewidth]{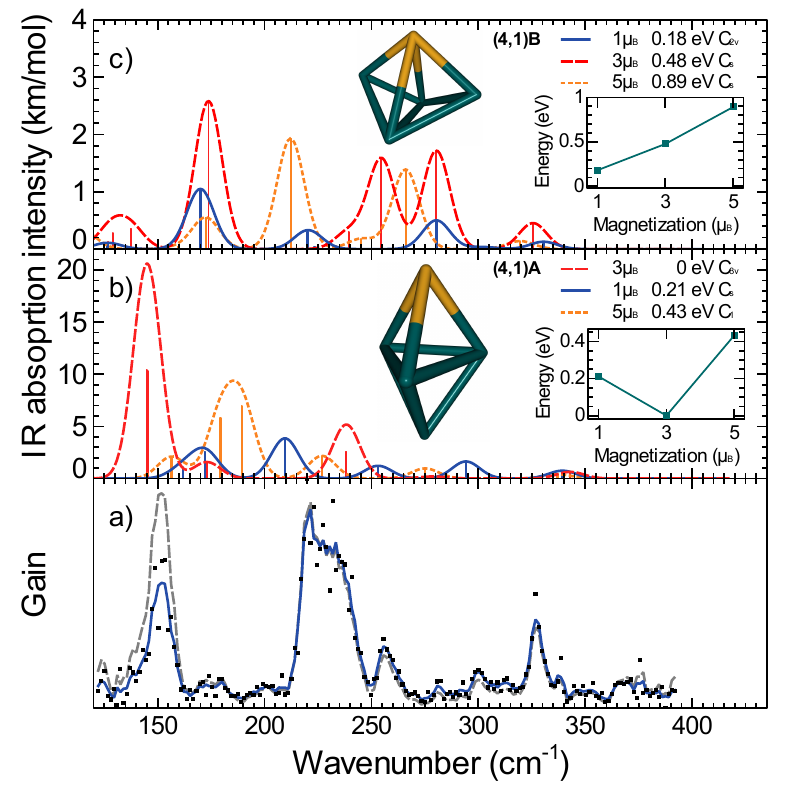}
		\caption{Experimental (panel (a)) and calculated ((b)-(d)) vibrational spectra of \ce{Nb4Co}. The insets show the energy as function of magnetization for each geometry. }
		\label{fignb4}
\end{figure}

In the experimental spectrum of \ce{Nb4Co} presented in Fig.~\ref{fignb4} (a), at least four modes can be distinguished, at 150, 230, 255 and 325~$\text{cm}^{-1}$. The three geometries lowest in energy are shown in Fig.~\ref{fignb4}(b)-(d). Geometry (4,1)A with $M=3$~$\mu_{\text{B}}$ is the lowest in energy and has $C_{3v}$ point group symmetry. 
Geometry (4,1)A consists of a triganol bi-pyramid, where \ce{Nb} and \ce{Co} are the axial atoms. In contrast, in geometry (4,1)B the \ce{Co} atom is part of the equatorial triangle. For geometry (4,1)B the $M=1$~$\mu_{\text{B}}$ state is the lowest in energy and is 0.18~eV higher compared to the lowest of (4,1)A. Note that both (4,1)A and (4,1)B are previously reported in Ref.~\onlinecite{Li2014}, where (4,1)A with $M=3$~$\mu_{\text{B}}$ was also found to be the lowest in energy. The vibrational spectrum of (4,1)A with $M=3$~$\mu_{\text{B}}$ consists of two large modes at 145 and 238~$\text{cm}^{-1}$ and smaller modes at 173, 278 and 342~$\text{cm}^{-1}$, and matches very well to the experimental spectrum. The vibrational spectrum of (4,1)A with $M=3$~$\mu_{\text{B}}$ is the only spectrum with two major modes around 150 and 230~$\text{cm}^{-1}$. Therefore, we assign geometry (4,1)A with $M=3$~$\mu_{\text{B}}$ to be the ground state of \ce{Nb4Co}.

\subsubsection{\ce{Nb5Co}}
\begin{figure}[ht]
		\centering
		\includegraphics[width=1.0\linewidth]{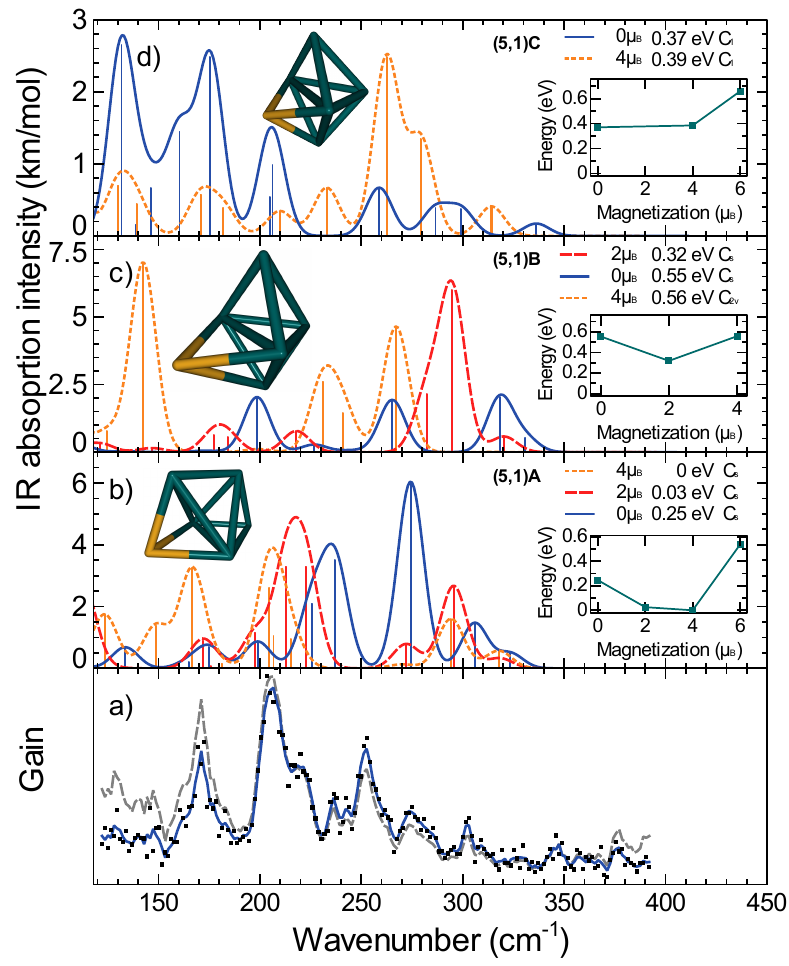}
		\caption{Experimental (panel (a)) and calculated ((b)-(d)) vibrational spectra of \ce{Nb5Co}. }
		\label{fignb5}
\end{figure}

In Fig.~\ref{fignb5} (b)-(d) the three geometries found to be lowest in energy for \ce{Nb5Co} are presented. Geometry (5,1)A consists of a dimer-capped rhombus with $C_{s}$ point group symmetry for all considered magnetic states and has been previously reported in Ref.~\onlinecite{Li2014} to be the lowest in energy for the $M=4$~$\mu_{\text{B}}$ state. We also find geometry (5,1)A in the $M=4$~$\mu_{\text{B}}$ state to be the lowest in energy, although the $M=2$~$\mu_{\text{B}}$ state is only 0.03~eV higher in energy. Geometries (5,1)B and (5,1)C both consist of a distorted \ce{Nb5} bi-pyramid with one of the faces of the bi-pyramid capped by the \ce{Co} atom. Geometries (5,1)B and (5,1)C differ in the distance of the \ce{Co} atom to the bi-pyramid. 
Whereas for the (5,1)B geometry the $M=2$~$\mu_{\text{B}}$ state is the lowest in energy, the (5,1)C geometry has a non-magnetic ground state which is 0.37~eV higher in energy compared to (5,1)A. The experimental spectrum of \ce{Nb5Co} in Fig.~\ref{fignb5} (a) shows three major bands at 170, 205 and 250~$\text{cm}^{-1}$, where the internal structure of the band at 205~$\text{cm}^{-1}$ indicates at least a second mode at 220~$\text{cm}^{-1}$. 
A smaller vibrational mode is present at 275~$\text{cm}^{-1}$. If the calculated spectra of Fig.~\ref{fignb5} (b)-(d) are compared to that of Fig.~\ref{fignb5} (a), both (5,1)
A $M=4$~$\mu_{\text{B}}$ and (5,1)C $M=0$~$\mu_{\text{B}}$ can only partially explain the experimental spectrum. Whereas (5,1) $M=4$~$\mu_{\text{B}}$ resembles the experimental spectrum below 230~$\text{cm}^{-1}$, the modes at 236 and 250~$\text{cm}^{-1}$ are not present in the calculated spectrum. Due to the similar vibrational spectrum and the low difference in energy between (5,1)A $M=2$~$\mu_{\text{B}}$ and (5,1)A $M=4$~$\mu_{\text{B}}$, the former can also not be excluded based on IR vibrational spectroscopy. The vibrational spectrum of (5,1)C $M=0$~$\mu_{\text{B}}$ agrees for the modes above 250~$\text{cm}^{-1}$, but deviates significantly in the relative IR absorption intensities between modes compared to the experimentally observed gain. Therefore, the IR gain spectrum of \ce{Nb5Co} might by due to the geometry (5,1)A with $M=2$~$\mu_{\text{B}}$ or $M=4$~$\mu_{\text{B}}$,  or geometry(5,1)C $M=0$~$\mu_{\text{B}}$. However, due to the finite temperature at which the experiment is performed, the vibrational spectrum might also be due to a combination of different geometries and magnetic states. On the other hand, the magnetic deflection experiments (see Section~\ref{magdefl}) were performed at a lower temperature than the vibrational experiments and strictly found Nb$_{5}$Co to be non-magnetic. Therefore, the (5,1)C geometry corresponding to the $M=0$~$\mu_{\text{B}}$ state is ascribed to be the ground state. 
	
\subsubsection{\ce{Nb6Co}}
\begin{figure}[b]
		\centering
		\includegraphics[width=1.0\linewidth]{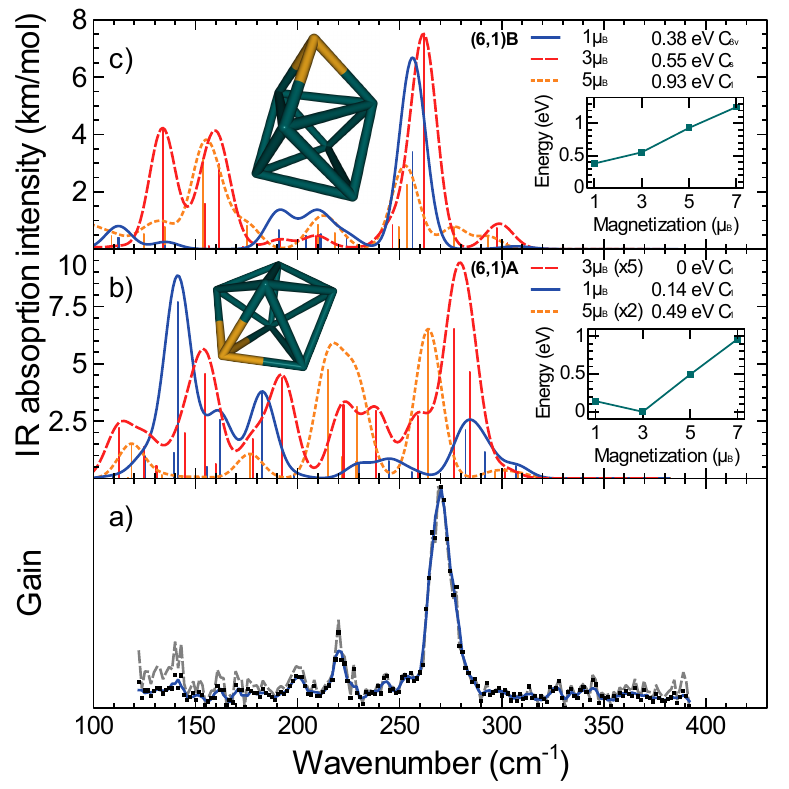}
		\caption{Experimental (panel (a)) and calculated ((b)-(d)) vibrational spectra of \ce{Nb6Co}. The IR absorption intensity of (6,1)A $M=3$~$\mu_{\text{B}}$ and $M=5$~$\mu_{\text{B}}$ are enhanced by a factor of 5 and 2 respectively to increase visibility. }
		\label{fignb6}
\end{figure}

The two geometries that were found to be the lowest in energy for \ce{Nb6Co} are shown in Fig.~\ref{fignb6} (b)-(c). Here geometry (6,1)A consists of a distorted pentagon with both sides capped with a single \ce{Nb} atom. Geometry (6,1)A in the $M=3$~$\mu_{\text{B}}$ state is obtained as the lowest in energy.
All magnetic states of the (3,1)A geometry have a $C_{1}$ point group symmetry. Geometry (6,1)B consists of two stacked \ce{Nb3} triangles, where the top triangle is capped with a \ce{Co} atom. For this geometry the $M=1$~$\mu_{\text{B}}$ state is the lowest in energy and has a $C_{3v}$ point group symmetry. The experimental IR spectrum of \ce{Nb6Co} is shown in Fig.~\ref{fignb6}(a) and contains a dominant mode at 270~$\text{cm}^{-1}$ and two smaller modes at 200 and 220~$\text{cm}^{-1}$. 
The vibrational spectrum of (6,1)B $M=1$~$\mu_{\text{B}}$ provides the best match to the experimental spectrum with a single dominant mode at 256~$\text{cm}^{-1}$ and several smaller modes constituting two bands at 190 and 210~$\text{cm}^{-1}$. In the vibrational spectrum of (6,1)A the bands at 220 and 264~$\text{cm}^{-1}$ have similar IR absorption intensities, which is in disagreement with the experimentally observed relative difference between these bands. All other geometries have significant vibrational modes below 190~$\text{cm}^{-1}$ where experimentally no modes are observed. Therefore, the (6,1)B geometry with the $M=1$~$\mu_{\text{B}}$ state is the ground state of \ce{Nb6Co}.

\subsubsection{\ce{Nb7Co}}
\begin{figure}[h]
		\centering
		\includegraphics[width=1.0\linewidth]{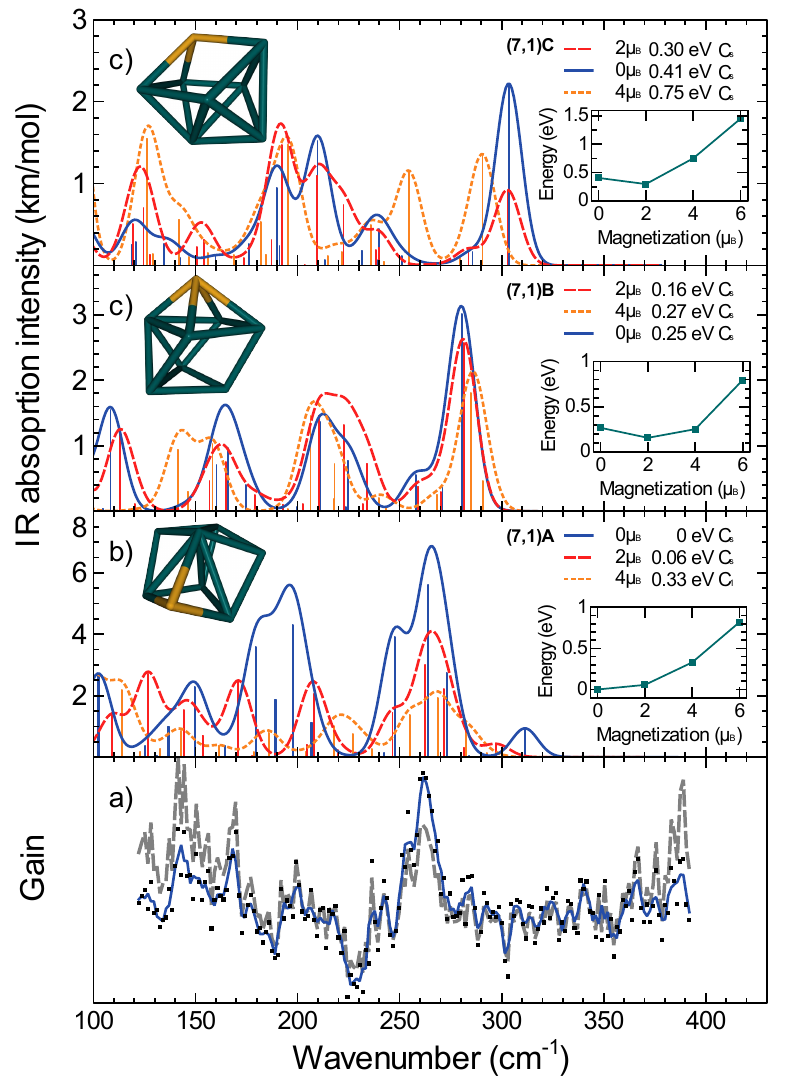}
		\caption{Experimental (panel (a)) and calculated ((b)-(d)) vibrational spectra of \ce{Nb7Co}. }
		\label{fignb7}
\end{figure}

The experimental spectrum of \ce{Nb7Co} in Fig.~\ref{fignb7} (a) shows a clear band at 260~$\text{cm}^{-1}$. The three geometries lowest in energy are shown in Fig.~\ref{fignb7} (b)-(d), where all geometries have either a symmetry plane or no symmetry at all. Geometry (7,1)A consists of a bicapped distorted \c{Nb} pentagon, where one of the faces is capped with a \ce{Co} atom.
The lowest magnetic state of geometry (7,1)A has a magnetic moment of $M=0$~$\mu_{\text{B}}$. This (7,1)A geometry has been previously reported in Ref.~\onlinecite{Li2014} to be also the lowest in energy. The (7,1)B geometry is formed by a \ce{Nb4} square capped on one side by a \ce{Co} atom and the other side by a \ce{Nb3} triangle. Geometry (7,1)C is described by a bipyramid containing a \ce{Co}atom at one of the tops and a single face of each pyramid is capped by a \ce{Nb} atom. In contrast to geometry (7,1)A, the magnetic ground states of geometries (7,1)B and (7,1)C are magnetic with $M=2$~$\mu_{\text{B}}$. Due to the reduction in symmetry, the vibrational spectrum of all geometries contain many vibrational modes.
The single experimental band at 260~$\text{cm}^{-1}$ can be both explained by (7,1)A $M=0$~$\mu_{\text{B}}$ and (7,1)B. However, the non-magnetic ground state of geometry (7,1)B is in very good agreement with experiment. Therefore, this geometry is assigned to be the ground state. 
	
\subsubsection{\ce{Nb9Co}}
	\label{nb9co}
\begin{figure}[htb]
		\centering
		\includegraphics[width=1.0\linewidth]{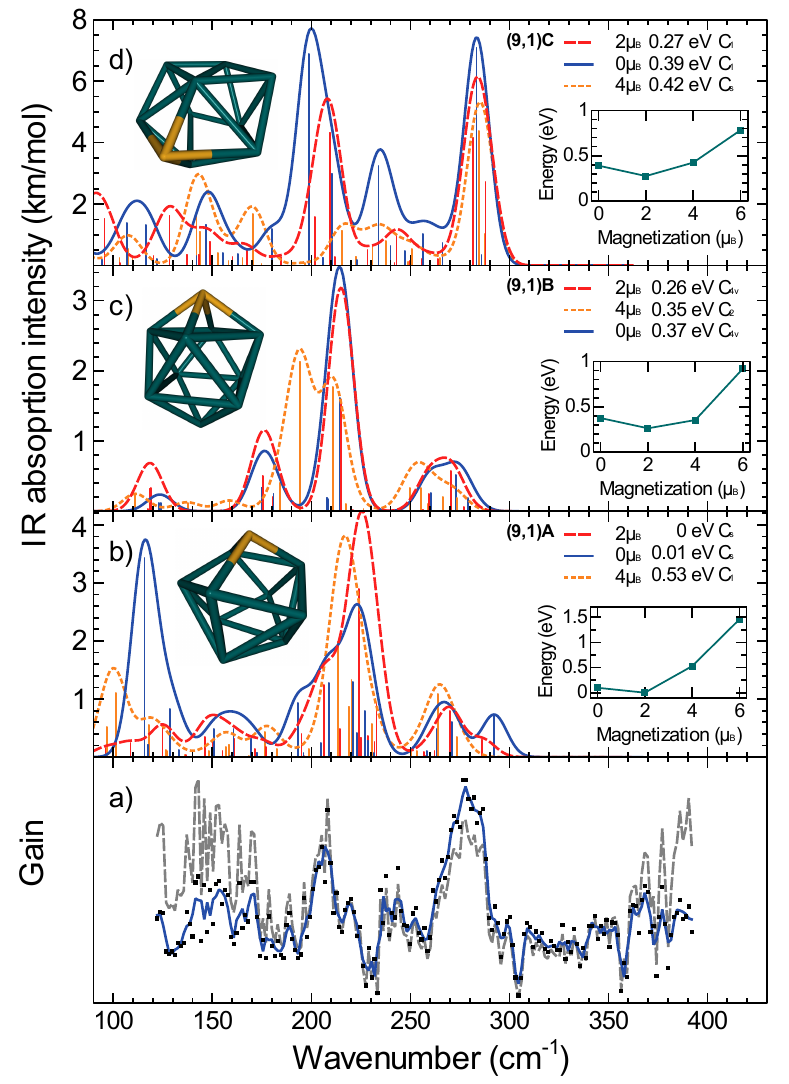}
		\caption{Experimental (panel (a)) and calculated ((b)-(d)) vibrational spectra of \ce{Nb9Co}. }
		\label{fignb9}
\end{figure}

	Fig.~\ref{fignb9} (a) shows the IR gain spectrum of \ce{Nb9Co}. Although this figure is not very well resolved, at least bands at 205, 240 and 280~$\text{cm}^{-1}$ can be identified. In Fig.~\ref{fignb9} (b)-(d) the three \ce{Nb9Co} geometries that were found to be the lowest in energy are presented. 
Here geometry (9,1)A consists of a \ce{Nb4} rhombus stacked with a \ce{Nb5} pentagon capped by a \ce{Co} atom. Note that geometry (9,1)A is distorted such that only a mirror plane symmetry remains. The geometry indicated by (9,1)B consists of two stacked \ce{Nb4} squares, where the two open faces are capped by a \ce{Nb} and \ce{Co} atom. The (9,1)C geometry is best described (yet poorly) by a distorted hexagon with a \ce{Nb} in the center and a \ce{Co} atom occupying a corner, and capped by a \ce{Nb3} triangle. Here for the geometry (9,1)C in the states $M=2$ and $M=0$~$\mu_{\text{B}}$ there is no symmetry, while in the $M=4$~$\mu_{\text{B}}$ state there is only a mirror plane. For geometry (9,1)C the $M=2$~$\mu_{\text{B}}$ state is found to be the lowest in energy, while the $M=0$~$\mu_{\text{B}}$ state is 0.12~eV higher in energy. 
If the calculated vibrational spectra of Fig.~\ref{fignb9} (b)-(d) are compared to the experimental spectrum, geometry (9,1)C with $M=2$ and $M=0$ provide the best match with dominant bands around 205 and 285~$\text{cm}^{-1}$ and an intermediate mode in-between. Therefore, the ground state of the \ce{Nb9Co} cluster is described by the (9,1)C geometry.

\subsection{Comparison with magnetic deflection results}
It is interesting to compare the magnetic moments obtained from the magnetic deflection experiments described in Section~\ref{magdefl} with those obtained above from an inspection of the vibrational spectra. In Table~\ref{tabcom} the second column contains the magnetic moments obtained from the best match of the calculated DFT vibrational spectra compared to experiment. For some clusters multiple magnetic moments are given, because for them it was not clear which vibrational spectrum matches the best with experiment. The third column corresponds to the magnetic moments observed in the magnetic deflection experiments (see Fig.~\ref{figdefl}).  
	
Except for Nb$_{5}$Co it appears that the magnetic moments predicted by the magnetic deflection experiments are substantionally larger. Part of this difference is due to not taking into account the orbital contribution to the magnetic moment within the DFT calculations. However, even if we would have considered them, it is well known that orbital moments can be highly underestimated in DFT especially for clusters.~\cite{larsco} Another possible reason for the difference in magnetic moments observed in Table~\ref{tabcom}, is an underestimation of the spin contribution within DFT. 
	
Unfortunately, for Nb$_{5}$Co and Nb$_{9}$Co we cannot be conclusive about the magnetic moment obtained from an inspection of the vibrational spectra. For Nb$_{5}$Co the zero magnetic moment would be in agreement with the magnetic deflection experiment, but this state is 0.37~eV higher in energy than the calculated ground state. Note that for Nb$_{6}$Co and Nb$_{7}$Co the best match of the calculated spectrum with experiment was also for a state higher in energy than the ground state, respectively 0.38 and 0.16~eV. On the other hand for Nb$_{3}$Co and Nb$_{4}$Co the spectrum calculated for the ground state provided the best match with experiment. For Nb$_{9}$Co the state with a magnetic moment of 2~$\mu_{B}$ would be the closest to the result of the magnetic deflection experiment. Here the state with a moment of 2~$\mu_{B}$ is 0.27~eV higher in energy than the ground state. 
	
\begin{table}[!ht]
		\begin{center}
			\caption{Here the second column corresponds to the magnetic moments obtained from the best match of the calculated DFT vibrational spectra with respect to experiment. The third column contains the magnetic moments obtained from the magnetic deflection experiments presented in Section~\ref{magdefl}.  }
			\begin{ruledtabular}
				\begin{tabular}{lcc}
					Cluster & M$_{vib}$ ($\mu_{B}$) & M$_{exp}$ ($\mu_{B}$)  \\ \hline
					Nb$_{3}$Co      &  2 & 6 \\
					Nb$_{4}$Co      &  3 & 6 \\
					Nb$_{5}$Co      & 0,2,4 & 0\\
					Nb$_{6}$Co     &  1 & 6 \\
					Nb$_{7}$Co     &  0 & 0\\
					Nb$_{9}$Co     & 0,2 & 3\\
				\end{tabular}
				\label{tabcom}
			\end{ruledtabular}
		\end{center}
\end{table}

\section{Theoretical investigation based on the Anderson impurity model} 
\label{theoraim}
In this section the physical origin is explained of the magnetic behavior obtained from the magnetic deflection experiments presented in Section~\ref{magdefl}. For example, it will be understood why some clusters are strongly magnetic, while others are non-magnetic. For this purpose an analyses based on the Anderson impurity is performed, where the ground state geometries obtained in Section~\ref{secvib} are required as an input.  
	
\subsection{Theoretical background}
There are two possible explanations for some Nb$_{x}$Co clusters being non-magnetic. It can be non-magnetic, because interactions of the Co atom with the Nb$_{x}$ host destroy the local moment at the Co site. More precisely, there is a competition between Jahn-Teller distortion working against the formation of a magnetic moment and the exchange interaction between Nb and Co preferring the existence of a magnetic moment. Another possibility is that the local moment at the Co site is screened by the delocalized electrons in the cluster, i.e. the Kondo effect. For both mechanisms it is crucial to understand physically when a local moment is formed on the Co site. In case of a magnetic (transition-metal) impurity resolved in a metallic non-magnetic host this is well established within the celebrated Anderson impurity model,
	
\begin{equation}
		\begin{gathered}
			H=\sum_{k,\sigma}\epsilon_{k\sigma}c_{k\sigma}^{\dagger}c_{k\sigma}\\
			+\sum_{\sigma}E_{d\sigma}d_{\sigma}^{\dagger}d_{\sigma}+U n_{d\uparrow}n_{d\downarrow}\\
			+\sum_{k,\sigma}V\Big(d_{\sigma}^{\dagger}c_{k\sigma}+c_{k\sigma}^{\dagger}d_{\sigma}\Big).
			\label{eqAIM}
		\end{gathered}
\end{equation} 
\newline
Here $E_{d\sigma}$ is the single-particle impurity energy level and $U$ is the onsite Coulomb repulsion between the impurity states. Further, the dispersion of the non-interacting electronic bath is given by $\epsilon_{k\sigma}$. The coupling between the impurity and bath states is described by $V$. Within this model the formation of a local moment depends on a delicate interplay between the onsite Coulomb interaction, the coupling strength between the impurity and bath states, the position of the bare impurity level (or equivalently the filling) and the positions of the bath energy levels (the dispersion). 
Within the static mean-field approximation the criterion for a local moment to exist is $U/\Gamma>\pi$. Here $2\Gamma=\pi V^{2} \rho(E_{F})$ is the effective hybridization, i.e. broadening of the impurity $E_{d}$ level, where $\rho(E_{F})$ is the density of impurity states at the Fermi level. From this criterion it is clear that a large onsite Coulomb interaction and small coupling between the impurity and bath are favorable for a local moment to exist.  
	
It is well known that Kondo physics occurs for the model described by Eq.~\ref{eqTk} at half-filling and in the limit where the hybridization can be treated perturbatively. More precisely, it can be shown that in this regime the virtual spin-flip scatterings of the bath electrons against the local impurity moment are the dominant processes occuring in the system. At low enough temperatures, below the Kondo temperature, they start to screen the local moment. For half-filling and by treating the hybridization perturbatively, the Kondo temperature $T_{K}$ can be estimated via
\begin{equation}
		\begin{gathered}
			T_{L}=U\Bigg(\frac{\Gamma}{2U}\Bigg)^{1/2}\exp\Bigg[\frac{-\pi|E_{d}||E_{d}+U|}{2U\Gamma}\Bigg],
			\label{eqTk}
		\end{gathered}
\end{equation} 
where the Kondo temperature is equal to \mbox{$T_{K}=0.041T_{L}$}~\cite{Hewson1993} The Kondo effect for very small systems has been been the subject of study already for several decades, e.g. for quantum dots. Theoretically, the Kondo effect was predicted to take place in quantum dots.~\cite{Lee1988,Lee1993,Meir1994} A few years later experiments confirmed these predictions.~\cite{Kouwenhoven1998,Kastner1998}

Although less studied within the Anderson impurity model, the situation of a magnetic impurity resolved in a semi-conductor or equivalently a bulk host with a band gap has also been addressed~\cite{Lau2014,Haldane1976}. It has been demonstrated that a local magnetic moment on the impurity is stabilized by the introduction of a band gap. In more detail a local moment can be formed even when the criterion above is not satisfied. Furthermore, the magnitude of the local moment increases with increasing band gap.
	
In Ref.~\onlinecite{Lau2014} the investigation of the Anderson impurity model for an impurity in a gapped host is extended to the situation of a finite sized host. Interestingly, it was found that on average the local moment grows with increasing band gap (HOMO-LUMO gap). Here on average should be understood as the local moment averaged over a number of random configurations of the discrete host energy levels for a fixed band gap. Further, it has been shown that in the regimes, where $V\ll E_{g}$ or $V\gg E_{g}$, the magnitude of the local moment merely depends on the size of the band gap ($E_{g}$) and not on the exact positions of the discrete energy levels of the host. Namely, for $V\ll E_{g}$ the effect of the hybridization is small no matter what the exact arrangement of the host energy levels is, while for $V\gg E_{g}$ the impurity level hybridizes with all host levels anyway. However, for the regime in between, $V\sim E_{g}$, the local moment strongly depends on the exact positions of the host energy levels. In Ref.~\onlinecite{Lau2015} these findings were successfully used to interpret the experimentally observed magnetic moments of Au$_{x}$Cr clusters. For example, the trend of the Au$_{x}$ host band gap was found to exactly follow that of the magnetic moment of the Au$_{x}$Cr clusters.        
	
\subsection{Computational details}
In this work we performed for the Nb$_{x}$Co clusters an analyses based on the Anderson impurity model in the same spirit as in Ref.~\onlinecite{Lau2015}. For this purpose the density functional theory (DFT)~\cite{Kohn1964,Kohn1965} is employed within the full-potential linear muffin-tin orbital method~\cite{Wills2010}. The local density approximation (LDA) exchange-correlation functional is used in the formulation of Perdew and Wang~\cite{Perdew1992}. 
For the Nb atoms the main valence basis functions were 4d, 5s and 5p states, while 4s and 4p states were treated as pseudo-core in a second energy set~\cite{Wills2010}. In case of Co, the 3s and 3p states were treated as pseudo-core, and the 3d, 4s and 4p states as the main valence states. In all calculations the valence states were treated scalar relativistically (without spin-orbit coupling). Since the employed DFT code works in {\bf k}-space, a supercell approach was used. A large unit cell of at least 14-\AA{ } dimensions was used in order to prevent the interaction between clusters of different unit cells. In these calculations the $\Gamma$ point was the only {\bf k}-point considered. The geometry of the clusters is obtained from the comparison of the experimental and DFT vibrational spectra performed in Section~\ref{secvib}. More precisely, the ground state geometries (3,1)A M=3, (4,1)A M=3, (5,1)C M=0, (6,1)B M=1, (7,1)B M=0 and (9,1)B M=2 are taken. Note that for Nb$_{9}$Co the structure with C$_{4v}$ symmetry is chosen. Namely for a magnetic cluster the Jahn-Teller distortion should be counteracted by the exchange interaction between Nb and Co.  
	
The effective onsite Coulomb repulsion $U$ between the 3d electrons of the Co impurity is obtained from DFT calculations in conjunction with the random phase approximation (RPA) within the full-potential linearized augmented plane wave (FLAPW) method~\cite{lapw}. All these calculations are performed with the GGA functional as formulated by Perdew, Burke and Ernzerhof~\cite{Perdew1996}. Here a large unit cell of at least 12-\AA{ } dimensions is used and also only the $\Gamma$ point is considered. Further, the plane wave cuttoff is 4.0 Bohr$^{-1}$. The actual RPA calculations are performed with the SPEX code, which uses the DFT calculations as an input~\cite{Friedrich2010}. The SPEX code uses the Wannier90 library to construct the maximally localized Wannier functions~\cite{Mostofi2008,Freimuth2008}. For this construction five 3d states and one 4s state are used for the Co atom.

\subsection{Results: Anderson impurity model}
Table~\ref{tabaim} presents for each Nb$_{x}$Co cluster its characteristic parameters related to the Anderson impurity model. The center of gravity of the Co 3d projected density of states $E_{d}$ and its weighted standard deviation $\Gamma$ are shown. Also are shown, the band gap (HOMO-LUMO gap) $E_{g}$ of the bare Nb$_{x}$ host for the geometry it has in the full Nb$_{x}$Co cluster and the effective onsite Coulomb interaction $U$ between the Co 3d electrons. Although Eq.~\ref{eqTk} is strictly speaking only valid for an impurity in a non-magnetic metallic host at half-filling in the limit of small hybridization, we employed it to obtain a rough estimate of the Kondo temperature $T_{K}$ for the Nb$_{x}$Co clusters. For convenience also the experimentally observed magnetic moment (see Fig.~\ref{figdefl} (d)) is presented in the last column. As can be observed, the impurity energy level $E_{d}$ and its broadening $2\Gamma$ are more or less constant as a function of cluster size. On the other hand, the band gap of the bare Nb$_{x}$ host strongly fluctuates as function of cluster size, while the effective onsite Coulomb repulsion slowly decreases as function of cluster size. 
	
\begin{table}[!ht]
		\begin{center}
			\caption{The Co impurity energy level $E_{d}$, broadening of the impurity level $2\Gamma$, energy gap $E_{g}$ (HOMO-LUMO gap) of the bare Nb$_{x}$ host and the effective onsite Coulomb interaction $U$ between the Co impurity 3d electrons within RPA for different Nb$_{x}$Co clusters. The sixth column contains a rough estimate of the Kondo temperature $T_{K}$ obtained from Eq.~\ref{eqTk}. For convenience also the experimentally observed total magnetic moment in $\mu_{B}$ is presented in the last column. }
			\begin{ruledtabular}
				\begin{tabular}{lcccccc}
					Cluster & $E_{d}$ (eV) & $\Gamma$ (eV) & $E_{g}$ (eV) & $U$ (eV) & $T_{K}$ (K)& M ($\mu_{B}$)  \\ \hline
					Nb$_{3}$Co      & -0.88     & 0.34 &  0.03  & 5.5 & 151 & 6 \\
					Nb$_{4}$Co      & -0.97     & 0.35 &  1.04  & 5.0 & 133& 6 \\
					Nb$_{5}$Co      & -1.28     & 0.35 &  0.11  & 4.6 & 68& 0\\
					Nb$_{6}$Co     & -1.16     & 0.34 &  0.002  & 4.3 & 81& 6 \\
					Nb$_{7}$Co     & -0.99     & 0.26 &  0.36  & 4.1   & 37& 0\\
					Nb$_{9}$Co     & -1.42     & 0.33 &  0.02  & 3.8   & 55& 3\\
				\end{tabular}
				\label{tabaim}
			\end{ruledtabular}
		\end{center}
\end{table}
	
As naively expected from Refs.~\onlinecite{Lau2014,Lau2015}, the magnitude of the local Co moment should follow the trend of the band gap of the isolated host as a function of cluster size. In other words a small band gap is expected for the clusters with zero magnetic moment, while a larger band gap is expected for the magnetic clusters. It is clear that this expectation is not verified by the results in Table~\ref{tabaim}. For example, magnetic Nb$_{3}$Co and Nb$_{6}$Co have a very small band gap compared with the non-magnetic Nb$_{5}$Co and  Nb$_{7}$Co clusters. 
	
It is also interesting to have an inspection of the criterion for the existence of a local moment in the Anderson impurity model. In case of an impurity with degenerate orbitals the criterion stated above is slightly modified into $(U+4J)/\Gamma>\pi$, where $J$ is the Hund exchange coupling between the impurity electrons. Even when the contribution of $J$ is neglected, it is clear from Table~\ref{tabaim} that the criterion is satisfied for all clusters. It was already known from Ref.~\onlinecite{Lau2015} that a magnetic impurity moment can occur even when the criterion above is not satisfied. However, it appears that the other way around is also possible, i.e. there is no magnetic moment even when the criterion is satisfied. 
	
Only considering the band gap of the bare host did not provide an explanation for some Nb$_{x}$Co clusters being magnetic and others non-magnetic. On the other hand for Au$_{x}$Cr it perfectly predicted the magnetic moment as function of cluster size. The reason is that the Au$_{x}$ host is inert, i.e. there is only a small coupling between the Cr impurity states and Au$_{x}$ host states. Therefore, Au$_{x}$ clusters can be considered to be in the regime $V\ll E_{g}$, where the size of the local moment solely depends on the band gap of the host and not on the exact positions of its energy levels. This is also apparent from the observation that the local moment of the Cr impurity is barely quenched in the Au$_{x}$Cr clusters. Contrary for the Nb$_{x}$Co clusters the magnetic moment strongly fluctuates as function of cluster size, which hints in the direction that we are in the regime $V\sim E_{g}$. Unfortunately, this cannot be directly verified from the parameters presented in Table~\ref{tabaim}. Namely, $\Gamma$ corresponds to the effective hybridization in which both $V$ and the density of states of the host are involved. However, indirectly one could argue that the Nb$_{x}$Co clusters are in the $V\sim E_{g}$ regime. From Ref.~\onlinecite{Lau2014} it is know that for $V\gg E_{g}$ the impurity moment is almost completely quenched, while for $V\ll E_{g}$ the moment should follow the size of the band gap. Since neither of the two is in agreemen with the results of Table~\ref{tabaim}, it is expected that the Nb$_{x}$Co clusters are in the $V\sim E_{g}$ regime. 
	
In the $V\sim E_{g}$ regime the exact positions of the host energy levels are known to be important. It would be helpful to be a bit more specific and to have a feeling for which host energy levels are important. For example, intuitively one would expect only host states within a range of about $V$ around the Fermi level (chemical potential) to be important. 
	
In order to verify this expectation we investigated the Anderson impurity model for an impurity with a single orbital coupled to 6 spin degenerate bath states. The impurity energy level and onsite Coulomb repulsion were chosen such that the single and double occupied isolated impurity states are symmetric around the chemical potential, e.g. $E_{d}=-1$ and $U=3$. Further, a total occupation (impurity plus bath) of 7 electrons was considered. The Anderson impurity model was solved exactly via exact diagonalization. Note that in Ref.~\onlinecite{Lau2014} a tight binding approximation was employed.   
	
In Table~\ref{tabaim2} the influence of different arrangements of 3 occupied and 3 unoccupied (occupied and unoccupied refers to the bare bath situation) spin degenerate host states on the impurity magnetic moment is presented. For all calculations $V=0.1$ is taken. The columns 2 to 7 correspond to the positions of the spin degenerate occupied and unoccupied host states, column 8 contains the band gap and the last column the magnetic moment on the impurity. From this table it is clear that indeed only host states within a range of $V$ are important in terms of the magnitude of the impurity magnetic moment. For example, a comparison of the first 5 calculations shows this. Also a comparison of the calculations 3, 8, 9 and 10 clearly indicates this. Another (trivial) observation can be made from calculations 4, 6 and 7. For these calculations the band gap is the same and the only difference is in the positions of the HOMO and LUMO levels with respect to the chemical potential. It appears that these exact positions are unimportant as long as the band gap is fixed. Finally, from calculations 3, 8, 9 and 10 it can also be concluded that not only the band gap itself, but also the number of states (density of states) involved is important.  
	
\begin{table}[!ht]
		\begin{center}
			\caption{The impurity magnetic moment (last column) for different arrangements of the occupied (columns 2 to 4) and unoccupied (columns 5 to 7) spin degenerate host states. The column with $E_{g}$ contains the band gap (HOMO-LUMO gap). }
			\begin{ruledtabular}
				\begin{tabular}{ccccccccc}
					& $E_{occ1}$ & $E_{occ2}$ & $E_{occ3}$ & $E_{unocc1}$ & $E_{unocc2}$ & $E_{unocc3}$ & $E_{g}$ & $M_{imp}$  \\ \hline
					1 & -0.5      & -0.5     & -0.5 &  0.5  & 0.5  & 0.5 & 1.0 & 0.98 \\
					2 & -0.3      & -0.3     & -0.3 &  0.3  & 0.3  & 0.3 & 0.6 & 0.97 \\
					3 & -0.1      & -0.1     & -0.1 &  0.1  & 0.1  & 0.1 & 0.2 & 0.78 \\
					4 & -0.05      & -0.05     & -0.05 &  0.05  & 0.05  & 0.05 & 0.1 & 0.22 \\
					5 & 0      & 0     & 0 &  0  & 0  & 0 & 0 & 0 \\
					6 & -0.1      & -0.1     & -0.1 &  0  & 0  & 0 & 0.1 & 0.21 \\
					7 & 0      & 0   & 0 &  0.1  & 0.1  & 0.1 & 0.1 & 0.23 \\
					8 & -0.5      & -0.1     & -0.1 &  0.1  & 0.1  & 0.5 & 0.2 & 0.89 \\
					9 & -0.3      & -0.1     & -0.1 &  0.1  & 0.1  & 0.3 & 0.2 & 0.88 \\
					10 & -0.2      & -0.1     & -0.1 &  0.1  & 0.1  & 0.2 & 0.2 & 0.86 \\
				\end{tabular}
				\label{tabaim2}
			\end{ruledtabular}
		\end{center}
\end{table}
	
	
Since the coupling strength $V$, the band gap and host density of states are important for the impurity magnetic moment, it would be natural to study the hybridization function corresponding to the Co 3d electrons. Namely, the imaginary part of the hybridization function is proportional to the coupling strength $V$ squared and the host density of states. Furthermore, in the regime $V\sim E_{g}$ the influence of the coupling of the impurity with the host cannot be considered as a (small) perturbation like in Au$_{x}$Cr. This coupling is already taken into account explicitly within the hybridization function.   
	
For details on how the hybridization function projected on the Co 3d states is obtained, the reader is referred to Ref.~\cite{granas2012} In short the Nb$_{x}$Co cluster is first calculated self-consistently within DFT. Then, from the obtained Kohn-Sham eigenstates and energies, the corresponding Green's function is constructed. Next, this Green's function is projected on the 3d states. This projected Green's function $G_{mm'}(E)$ and the hybridization function of the Co 3d states $\Delta_{mm'}(E)$ are related by
	
\begin{equation}
		\begin{gathered}
			G_{mm'}(E)=\Bigg[ E - \epsilon_{mm'}+\mu - \Delta_{mm'}(\omega)\Bigg], \\
			\text{with  }\Delta_{mm'}(E)=\sum_{k} \frac{V^{*}_{km}V_{km'}}{E-\epsilon_{k}+\mu}.
			\label{eqhyb}
		\end{gathered}
\end{equation} 
Here, $E$ is the energy, $V_{km}$ represent the coupling strength of the impurity state $m$ with bath (host) state $k$, $\epsilon_{mm'}$ is obtained from the local projection of the DFT Kohn-Sham Hamiltonian and $\epsilon_{k}$ corresponds to the energies of the bath states. From the expression of the hybridization function in terms of the coupling strengths and bath energy levels, it is clear that different choices of them can lead to the same hybridization function and thus Anderson impurity problem. Therefore, unless the $V_{km}$ matrix elements are computed directly, it is hard to explicitly determine whether Nb$_{x}$Co corresponds to the $V\sim E_{g}$ regime. However, this determination is not necessary to understand the physical origin of the presence or absence of magnetism in the Nb$_{x}$Co clusters.   
\begin{figure}[tb]
		\centering
		\includegraphics[width=1.0\linewidth]{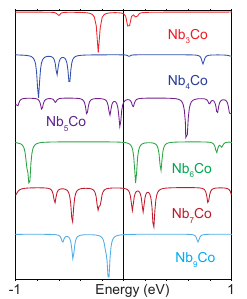}
		\caption{The imaginary part of the hybridization function for the Co 3d electrons for the different Nb$_{x}$Co clusters.}
		\label{fighyb}
\end{figure}
From the discussions above we know that the HOMO-LUMO gap, the density of states at the HOMO and LUMO levels, the coupling $V$ between the impurity and host states, and the onsite Coulomb repulsion $U$ are important for the impurity magnetic moment. The first three are captured by the (imaginary part of the) hybridization function. Therefore, in Fig.~\ref{fighyb} the imaginary part of the total (trace of $\Delta_{mm'}(E)$) hybridization function for the Co 3d states is shown for the different Nb$_{x}$Co clusters. From this figure an estimate
can be made of the coupling strength $V$. Assuming that the peak of Nb$_{3}$Co at -0.25~eV is due to the coupling with only one bath state, would require a $V$ of about 0.37~eV. Therefore, the hybridization function is only plotted roughly in this range around the chemical potential (zero energy). 
	
From the model calculations presented in Table~\ref{tabaim2} it is expected that a small HOMO-LUMO gap and large hybridization around the HOMO and LUMO levels is unfavourable for a magnetic moment. A discussion solely based on the hybridization functions of Fig.~\ref{fighyb} is complicated by the fact that the onsite Coulomb repulsion is not constant over the range of clusters investigated. However, for two clusters differing only by one Nb atom in size the difference in the onsite Coulomb interaction is small. Therefore, in the following the hybridization functions will be compared cluster for cluster. From Fig.~\ref{fighyb} it appears that Nb$_{3}$Co has a much stronger hybridization around the chemical potential (zero energy) than Nb$_{4}$Co. More precisely for Nb$_{3}$Co there is a peak at about -0.25~eV and 0.1~eV, while Nb$_{4}$Co has a peak at about -0.5~eV and a very tiny one at 0.05~eV. Since the gap between the peaks is larger and the total height of the peaks is smaller for Nb$_{4}$Co, a larger magnetic moment is expected for Nb$_{4}$Co compared to Nb$_{3}$Co. This is comfirmed by the magnetic deflection experiment (see last column of Table~\ref{tabaim} and Fig.~\ref{figdefl}). 
	
By going from magnetic Nb$_{4}$Co to non-magnetic Nb$_{5}$Co, it is clear that there is a huge increase of hybridization around the chemical potential. Therefore, in addition with a smaller onsite Coulomb interaction it is indeed expected that Nb$_{5}$Co has a much smaller tendency to be magnetic than Nb$_{4}$Co (and Nb$_{3}$Co). Then, by going from non-magnetic Nb$_{5}$Co to magnetic Nb$_{6}$Co, there is a huge decrease of hybridization around the chemical potential. More precisely, there is a huge increase from about 0.15~eV to 1.0~eV in the separation between the first peak below and above the chemical potential. Thus, in accordance with experiment Nb$_{6}$Co is expected to have a larger tendency to be magnetic than Nb$_{5}$Co. 
Next, magnetic Nb$_{6}$Co and non-magnetic Nb$_{7}$Co will be compared. As expected the hybridization around the chemical potential is larger for Nb$_{7}$Co than for Nb$_{6}$Co. Interestingly, Nb$_{7}$Co has a similar hybridization around the chemical potential as Nb$_{3}$Co. However, Nb$_{3}$Co has an onsite Coulomb interaction which is 1.4~eV larger than for Nb$_{7}$Co. Finally, non-magnetic Nb$_{7}$Co and magnetic Nb$_{9}$Co are compared. Although Nb$_{9}$Co has a quite large peak at about -0.15~eV, the difference between the first peak below and above the chemical potential is much larger. Therefore, the effective hybridization around the chemical potential is as expected smaller for Nb$_{9}$Co than for Nb$_{7}$Co. To conclude, for Nb$_{3}$Co to Nb$_{7}$Co and Nb$_{9}$Co the effective hybridization around the chemical potential is in agreement with the experimentally observed magnetic behavior. 
	
	
Above we performed an analysis based on the Anderson impurity model in order to explain the experimentally observed magnetic behavior. From an inspection of the hybridization function and the onsite Coulomb repulsion a trend in agreement with experiment could be predicted. However, based on these observations it cannot be explained whether the moment is (completely) quenched or Kondo screened. Therefore, we made an estimate of the Kondo temperature for the clusters from Eq.~\ref{eqTk}, which are presented in the sixth column of Table~\ref{tabaim}. In case the non-magnetic clusters occur due to  a complete Kondo screening, higher Kondo temperatures are expected for the non-magnetic clusters than for the magnetic clusters. From Table~\ref{tabaim} it can be observed that the results are not in accordance with this expectation. For example, the highest Kondo temperatures are observed for magnetic Nb$_{3}$Co and Nb$_{4}$Co. Further, non-magnetic Nb$_{5}$Co and Nb$_{7}$Co have a smaller Kondo temperature than magnetic Nb$_{6}$Co.  
	
In addition we searched for signatures of the Kondo effect in the Nb$_{x}$Co clusters from the experimental side. For this purpose the temperature dependence of the magnetic deflection experiments was investigated. In case of the Kondo effect it is expected that by approaching the Kondo temperature from below the screening of the local Co moment reduces. An inspection of Table~\ref{tabaim} shows that Nb$_{5}$Co has a Kondo temperature of 68K and Nb$_{7}$Co of 37K. However, even for temperatures up to 70K both clusters still appeared to be strictly non-magnetic. These results indeed indicate that the Kondo effect is not responsible for Nb$_{5}$Co and Nb$_{7}$Co to appear non-magnetic.

\section{Conclusion}
	\label{concl}
In this work we performed magnetic deflection experiments on Co doped Nb clusters from which we made the interesting observation that some clusters are strongly magnetic, while others are non-magnetic. Further, it appeared that the magnetic behavior of the clusters could be divided into two regimes. For Nb$_{x}$Co clusters with $x>=14$, the magnetic to non-magnetic behavior is exactly determined by having an odd or even number of atoms in the cluster, i.e. having an unpaired electron or not. Note that this behavior was also observed for pure Nb clusters. Then, in the region $x<14$ strong fluctuations in the magnetic moment as function of cluster size are observed in contradiction with the odd/even behavior described above.
	
There are in principle two possible explanations for some clusters being non-magnetic. Either the local moment at the Co site is completely quenched or it is screened by the delocalized electrons of the cluster, i.e. the Kondo effect. In order to reveal the physical origin, we conducted a combined theoretical and experimental investigation. 
	
First, we made a comparison of the experimental vibrational spectra with those obtained from a DFT study. This served two purposes. It provides the ground state geometry of the clusters. Further, due to the dependence of the vibrational spectrum on the magnetic moment, the performance of DFT in predicting the magnetic moments can be investigated. We found that not for all clusters  it could be determined which calculated vibrational spectrum has the best agreement with experiment. However, for those it could, we found that the DFT magnetic moments were considerably smaller than those obtained from the magnetic deflection experiments. This is due to a neglect of the orbital moments in our DFT calculations and underestimation of the spin moments within DFT. 
	
Second, with the obtained ground state structures as an input we performed an analyses based on the Anderson impurity model. It appears that the non-magnetic clusters are due to a complete quenching of the local Co moment and not due to the Kondo effect. In addition, the magnetic behavior of the Nb$_{x}$Co clusters can be understood from an inspection of their electronic structure. Here magnetism is favored when the effective hybridization around the chemical potential is small, while the absence of magnetism is signalled by a large effective hybridization around the chemical potential.

\section{Acknowledgments}
The Nederlandse Organisatie voor Wetenschappelijk Onderzoek (NWO) and SURFsara are acknowledged for the usage of the LISA supercomputer and their support. L.P. and M.I.K. acknowledges a support by European ResearchCouncil (ERC) Grant No. 338957.

\end{document}